\documentstyle[12pt]{article}

\def\frac#1#2{{\displaystyle#1\over\displaystyle#2}}

\def\pdr{\partial}

\begin{document}

\begin{center}
{\Large Do we need to recourse to Amp\`{e}re-Neumann electrodynamics to explain
wire fragmentation in the solid state?}

\vspace{3mm}

{A. Lukyanov, S. Molokov}

\vspace{3mm}

\noindent \hspace*{-0.4cm}{Coventry University, School of Mathematical and
Information Sciences, Priory Street, Coventry CV1 5FB, U.K.}
\end{center}

\baselineskip=8mm

\begin{abstract}
{\small 
Exploding wires are widely used in many experimental setups and 
pulsed power systems. However, many aspects of the process of wire
fragmentation still remain unclear. If the current density is not too
high, the wire may break up in the solid state. The experiments have
shown that the wires break in tension due to longitudinal forces of
unknown nature. 

In a series of papers, see 
\cite{Graneau84a,Graneau84b,Graneau87,Graneau94} Graneau argued that neither mechanical 
vibrations induced by the electromagnetic pinch force nor thermal expansion could 
have been responsible for the wire disintegration because they were too weak.
To explain the phenomenon he appealed to the obsolete Amp\`{e}re force law 
as opposed to the conventional Biot-Savart force law. Graneau argued
that the Amp\`{e}re force law would lead to a longitudinal tension in the
wire, although his calculations may have been in error on this point.
Therefore, Graneau's explanations induced a controversy in
electrodynamics with a number of authors arguing {\it pro} and {\it con}.

Previous theoretical and numerical investigations served
to provide a search for these forces without recourse to unconventional
electrodynamics have identified the pinch effect
and thermal expansion as a source of strong longitudinal vibrations.
But the tensile stress component has been proved to be present only in the
wires having free ends. Thus the mechanism doesn't give a
satisfactory explanation of the phenomenon in the wires with clumped ends.
In this investigation, we use a simplified magneto-thermo-elastic model
to study flexural vibrations induced by high pulsed currents in wires
with clumped ends on account of their role in the disintegration process. 
Several aspects are studied, namely (i) the buckling instability due to
simultaneous action of the thermal expansion and the magnetic force, 
(ii) the flexural vibrations induced in initially bent wires.
It is shown that the induced flexural vibrations are strong enough 
to lead to the breaking of the wire in a wide range of parameters.}
\end{abstract}

\vspace*{0.3cm}

\section{Introduction}

The phenomenon of wire fragmentation in the solid state by high pulsed 
currents was studied
experimentally by Nasilowski \cite{Nasilov64}, and Graneau 
\cite{Graneau84a,Graneau84b,Graneau87}. They
observed that a sufficiently strong electric current would shatter a thin
metal wire under a broad range of conditions (various wire material and
geometry, different current types, etc.). As a result of the explosion the
wires fragmented into 2-100 pieces with apparent signs of longitudinal
tensile stress. The experimental results, as well as (sometimes
controversial) early attempts to explain the phenomenon have been reviewed
by Graneau \cite{Graneau94} , Hong \cite{Hong94} and Molokov\&Allen \cite{SMolokov}.

It is obvious that an electric current passing through a metal wire induces
stress waves. Their origins are in (i) the thermal expansion owing to the
volumetric Joule heating, and (ii) the Lorentz-, or pinch-, force.
These forces depend on the magnitude and the rate of change of the 
current passing through the wire. They grow as either of these parameters increases.
 
Ternan \cite{Ternan86} suggested that in Graneau's experiments with wires with free
ends standing stress waves may be induced as a result of thermal expansion. 
Using a simple one-dimensional model he showed that the
resulting stresses were sufficient to lead to a fracture.

This mechanism of fragmentation of wires with free ends has been explored in
more detail by Molokov\&Allen in \cite{SMolokov}. They employed a
magneto-thermo-elastic model of the stress-wave propagation, and solved the
problem numerically on assumption of axisymmetric nature of vibrations . 
Their conclusion was that the characteristic value of
both compressive and tensile longitudinal stresses obtained for the aluminium
wire with radius $a=0.6\,\mbox{mm}$ carrying current $I=5\,\mbox{kA}$ and
having free ends (the parameters relevant to the experiment in \cite{Graneau84b}) 
would be about $33\,\mbox{MPa}$ per unit length of the wire. The stress has
been shown to
grow linearly with the wire length and thus, for sufficiently long wires,
the resultant stress exceeds the ultimate strength of aluminium. 
This is $75\,\mbox{MPa}$ at $100^{\circ }\,\mbox{C}$ going down
to $17\,\mbox{MPa}$ at $320^{\circ }\,\mbox{C}$ \cite{Smithells}.

Although the mechanism does give large value of tensile stresses, it cannot
explain the fragmentation of wires with clamped ends (as in the experiments 
\cite{Nasilov64,Graneau87}) because within the axisymmetric model in this case 
the stresses can be only
compressive \cite{SMolokov}. It is clear, however, that in the wires with
clamped ends flexural vibrations may be induced. The apparent
signs of these modes have been observed in the Graneau's 
experiments on the wires with firmly clamped ends. The resultant wire
shape, after breaking, clearly indicated on that, see photographs in 
\cite{Graneau87}. 

In this paper we carry out a special study of flexural vibrations to account for 
their role in fragmentation of wires with clamped ends. 
In section 2, we present the formulation of the problem and discuss
simplifications of the model. In section 3, the linear stability of the
system is considered. In section 4, we carry out a qualitative analysis
of expected stresses. In section 5, numerical algorithm and
results are presented.

\section{Formulation}

\bigskip Consider a wire which is firmly clamped between two points located along the 
$z$-axis. Let the distance between these points be $L$.   
We restrict our attention to the case of plane motion, and denote
the deflection of the wire from the axis by $X$ which is a function of $z$ and time $t$. 
The deflection of the wire in the initial undeformed state is denoted by $X_0$.
Possible three-dimensional
modes, not considered here, may only increase stresses, especially at the
clamped ends. 

Our theoretical consideration is based on a simplified model
valid for sufficiently long or sufficiently thin wires. In this model the
deflection $X$ of the wire is assumed to be much lower
than the wire length, i.e. $|X|/L\ll 1$. 
Since large deflections of the order
of the wire length are not expected in the wire explosion experiments, this
restriction is not expected to be very stringent. Another small parameter
used in the model is the ratio $a/L<<1$, where $a$ is the wire radius. This condition
is usually well satisfied.
Plastic deformations
are not considered here although they may be important in the vicinity 
of the melting point. Thus all deformations are assumed 
to be elastic.

Then the deflection $X$ obeys the following equation (\cite{LL}): 
\begin{equation}
\label{One}
\rho S\frac{\partial^2 X}{\partial t^2}+E\frac{\partial^2}
{\partial z^2}J\frac{\partial^2 X}{\partial z^2}-
E\frac{\partial^2}
{\partial z^2}J\frac{\partial^2 X_0}{\partial z^2}
-G\frac{\partial^2 X}
{\partial
z^2}-F_{x}-\frac{\partial C_{y}}{\partial z}=0  
\end{equation}
where $\rho $ is the material density, $S$ is the wire cross-section which
may be a function of $z$, $J$ is the effective moment of inertia of the
wire cross-section (for a circular cross-section $J=\pi a^{4}/4,$
where $a$ is the wire radius), $E$ is the Young modulus, $G$ is the force
along the $z$ axis applied at the wire ends, $F_{x}$ is the distributed
external force per unit length along the $x$ direction and $C_{y}$ is the
distributed moment of force along the $y$ direction per unit length.

For the physical situation considered here, the force $G$ can be represented
as
\begin{center}
$G=ES\{(\tilde{l}-\tilde{l_0})/L-\alpha \lbrack T(t)-T_{0}\rbrack\},$
\end{center}
The second term is a compressive force due to the thermal expansion, $\alpha 
$ being the linear expansion coefficient. The first, nonlinear term is a
retarding force due to the wire deformation from the initial state, where
\begin{equation}
\tilde{l}-\tilde{l_0}=\frac{1}{2}{\displaystyle\int_{0}^{L}}
\left( \frac{\partial X}{\partial z}\right)^{2}- 
\left(\frac{\partial X_0}{\partial z}\right)^{2}dz  
\end{equation}
is the increment of the wire length due to the deflection. 
When $X=X_0$, $\tilde{l}=\tilde{l_0}$ and the retarding force equals to zero,
as required. 

The boundary conditions are those for the clamped ends, namely 
\begin{equation}
\label{clends}
X=\frac{\partial X}{\partial z}=0\qquad \mbox{at}\,\,z=0 
\,\, \mbox{and}\,\, z=L.
\end{equation}
The initial conditions are
\begin{equation}
\label{incond}
X=X_0 \quad t=0.
\end{equation}

\subsection{The magnetic force}

In long, thin wires under consideration $a/L<<1$, uniform current is directed
perpendicular to the cross-section. Thus any possible forces of magnetic 
nature act only in the cross-section plane. These forces create no moment and 
thus $C_y=0$ in our case.   

The net distributed force ${\bf F}$ can be calculated directly from
\begin{equation}
\label{theforce}
{\bf F}=\int{\bf j\times B}\, dS.
\end{equation}
The integral in (\ref{theforce}) is taken over the wire cross-section area.
The magnetic field is given by the Biot-Savart law 
\begin{equation}
\label{BS}
{\bf B}=\frac{\mu_0}{4\pi}\int\frac{\bf j\times R}{R^3}dV^{'},
\end{equation}
where $dV^{'}$ is the volume element on the wire,  
${\bf j}$ is the current density, and the vector ${\bf R}={\bf r}-{\bf r^{'}}$
is directed to the field point as usual.

To calculate the $x$ component of the force, $F_x$, one can split the integral
(\ref{BS}) over the wire length $z^{'}$ into two parts ${\bf B}={\bf B_{in}}+{\bf B_{ex}}$. 
The first part of the integral ${\bf B_{in}}$ is taken over
the range $|z^{'}-z|<\Delta$, where $a<<\Delta<<\xi_L$, $\xi_L$ being the
characteristic length scale of the function $X(z)$. The second part ${\bf B_{ex}}$ is
taken over the rest of the wire length $\Delta<|z^{'}-z|<L$, whereby the volume
integral reduces to the integration over the wire length only.
That is,
$$
{\bf B_{ex}}=\frac{\mu_0}{4\pi}\int\frac{\bf I\times R}{R^3}dl^{'},
$$
where $I$ is the total current.
The first integral can be evaluated in general with the
assumption of a uniform current distribution and in the approximation 
$a<<\xi_L$, $|X|/\xi_L<<1$.  The first inequality has been implicitly introduced
in fact when we set the length $\Delta$, the second inequality just represents
the fact that the wire is slightly deflected from a straight line.   
The first integration 
asymptotically gives the expression for the net force as follows
\begin{equation}
\label{force1}
\left\{ F_x\right\}_{in}=
-\frac{\mu_0 I^2}{4\pi}\frac{\partial^2 X}{\partial z^2}
\{\ln(2\Delta/a)-3/4\}+O((X/\xi_L)^2,\Delta/\xi_L,(a/\Delta)^2).
\end{equation}
That is the force is proportional to the local wire curvature \cite{Thompson64}.
The second part cannot be integrated in a general case and 
one needs to make certain assumptions about the wire form and thus 
about the function $X(z)$ itself. In the case of an infinitely long wire 
with periodic lateral perturbations 
$X=\widetilde{X}\cos(kz)$, 
$k=2\pi/\xi_L$, one gets
\begin{equation}
\label{force2}
\left\{ F_x\right\}_{ex}=
-\frac{\mu_0 I^2}{4\pi}\frac{\partial^2 X}{\partial z^2}
\{-C+1/2-\ln(k\Delta)\}+O(k\Delta,(X/\xi_L)^2,(a/\Delta)^2),
\end{equation}
where $C\simeq 0.577$ is the Eiler's constant and we have substituted
$X k^2=-\frac{\partial^2 X}{\partial z^2}$.
Combining two parts we obtain
\begin{equation}
\label{forcem}
F_x=-\frac{\mu_0 I^2}{4\pi}\frac{\partial^2 X}{\partial z^2}
\{\ln(2/ka)-C-1/4\}.
\end{equation}
Note, that the expression (\ref{forcem}) is identical to the formula
obtained in \cite{Moon75} in a particular case of small periodic deflections $|X|<<a$.

It is clear that the assumption of an infinitely long wire to calculate the force, 
$F_x$, is not  very rigorous,
since in experiments the circuit is always closed. This implies that
in addition to the magnetic force  
by the wire current itself, there may
be a component generated by the currents flowing 
in the external circuit. This effect can be taken into consideration
by substituting an appropriate circuit form, in other words 
the function $X(z)$ in the second integral.
 
Alternatively, this force can be obtained
by means of the virtual work principle. 
To estimate how strong the effect could be, we will use a general
expression for the inductance of a closed circuit
$\Lambda=\frac{\mu_0}{2\pi}L_s\ln(L_s/a_s)$, where $L_s$ is the circuit characteristic 
length and $a_s$
is the radius of the circuit wire, see \cite{LLEL}.
Then varying the circuit length $L_s$
one can calculate the subsequent variation of the associated magnetic
energy $E_m=\frac{1}{2}\Lambda I^2$. Then, equalising the net force
acting on the circuit $F_N$ with the variation of the energy we get
\begin{equation}
\label{avforce}
F_N\delta L_s=\frac{\mu_0}{4\pi} I^2 (\ln(L_s/a_s)+1)\delta L_s.
\end{equation}

An average force acting on a unit length is $f=F_N/L$. Then,
for the average force from (\ref{avforce}) we get
$f\simeq \frac{\mu}{4\pi}_0 I^2 (\ln(L_s/a_s)+1)/L_s$.
The first, logarithmic
part of the net force represents contribution from the end effects, see (\ref{force1}).
The second part gives the integrated force from the circuit as a whole. If $L<<L_s$, 
one can assume that the force is constant throughout
the wire length in the first approximation. Then,
this ''constant'' force acting on a wire of length $L$ will create 
longitudinal stress $\sigma_{\|}=\frac{\mu_0^2 I^4}{60480}
\frac{\ln(L_s/a_s)^2}{E(\pi a^2)^4}\frac{L^6}{L_s^2}$, see the problem 
in \cite{LL} on page 93. For a copper wire with $L=30\,\mbox{cm}$,
$a=0.6\,\mbox{mm}$ and the circuit with $I=5\,\mbox{kA}$, $L_s=20\,\mbox{m}$
and $a_s=1\,\mbox{cm}$ the stress is amounting to 
$\sigma_{\|}\simeq 8.4\,\mbox{MPa}$. 
This is only an estimation, and
the effect needs special analysis since the geometry 
of external circuits is usually unknown. Moreover, using a symmetrical
circuit with two loops from both sides of the wire one can completely
compensate the external magnetic force.

We leave this question for further
detailed investigation and neglect the contribution from 
the external circuit in our analysis. Further we will use formula (\ref{forcem}).

\subsection{The temperature behaviour}
The temperature behaviour with time is governed by the passing current. 
It can be calculated 
according to the direct Joule heating of the wire material from 
\begin{equation}
\label{therms1}
\rho c_v \frac{\pdr T}{\pdr t}=j^2/\sigma,
\end{equation}
neglecting by the process of thermal conductivity. This assumption would hold for
thermally isolated wires with uniform current distribution.
In expression (\ref{therms1}),
$j$ is the current density, $\rho$ and $\sigma$ are the density and 
the electrical conductivity of the metal, 
$c_v$ is the specific heat.
From the above equation using the well-known inverse dependence 
of the conductivity on temperature
$\sigma=\sigma_0\frac{T_0}{T}$
one can obtain
$$T=T_0\exp(\gamma_T\int_0^t f(t^{'})^2dt^{'}),$$ where 
$\gamma_T=\frac{j_0^2}{\sigma_0 T_0\rho c_v}$ is the characteristic temperature 
rise-time, $j_0$ is maximum current density. The function $f(t)$ 
defines current rise time profile. 

\section{Linear stability}

To investigate possible instabilities in the system, 
we first carry out a linear stability analysis 
assuming for a moment that all the parameters, such as 
current and wire temperature, are constants independent of time.
Even though, in a real situation, they vary quite fast with time and
numerical methods must be involved to solve the system, our simplified
analysis will form a basis for qualitative interpretation and
understanding of the results.

\subsection{Stability of an initially straight wire}
Consider an initially straight wire by letting $X_0=0$.
General analysis of the system (\ref{One}) shows that two basic types
of buckling instabilities may develop. They are due to thermal expansion,
expressed by the force $G$ applied at the wire ends \cite{LL}, and the magneto-elastic
buckling instability \cite{Leon52} expressed by force $F_x$. 
In both cases, the instability has a threshold character. When
either the force $G$ or the current $I$ exceed some
critical value, new stable states appear. The initially stable state 
$X=0$ then becomes unstable and buckling occurs. 

We are looking for non-trivial stationary solutions of the linearized system (\ref{One}),
(\ref{clends}), (\ref{incond}). They are given by the equation
\begin{equation}
\label{OneS}
EJX^{IV}-GX^{''}-F_{x}=0,  
\end{equation}
with the boundary conditions
\begin{equation}
\label{clendsS}
X=X^{'}=0\qquad \mbox{at}\,\,z=0 
\,\, \mbox{and}\,\, z=L.
\end{equation}
Here, the temperature, $T$, and the total current, $I$, are assumed to be  
independent of time, 
and so are the functions $G=-\alpha ES[T-T_0]$ and 
$F_x=-\frac{\mu_0 I^2}{4\pi}X^{''}\{\ln(2/ka)-C-1/4\}$.

To obtain a solution to (\ref{OneS}) it is convinient to use a complete set of orthogonal 
functions $\left\{ {\mathcal X}_i \right\},\,\, i\geq 1$ defined by the problem
\begin{equation}
\label{orts1}
\begin{array}{c}
{\mathcal X}_i^{IV}+\lambda_i^2 {\mathcal X}_i^{''}=0\\
{\mathcal X}^{'}_i(0)={\mathcal X}^{'}_i(L)={\mathcal X}_i(0)={\mathcal X}_i(L)=0
\end{array}
\end{equation}
with the orthogonality given by
\begin{equation}
\label{ortf}
\begin{array}{rl}
\int_0^L {\mathcal X}^{''}_i{\mathcal X}^{''}_k dz=&\left\{
\begin{array}{cl}
\, 0,\,\,&i\neq k\\
\,\frac{\lambda_i^4}{2}L,\,\,&i=k
\end{array}\right.\\
\\
\int_0^L {\mathcal X}^{'}_i{\mathcal X}^{'}_k dz=&\left\{
\begin{array}{cl}
\, 0,\,\,&i\neq k\\
\,\frac{\lambda_i^2}{2}L,\,\,&i=k
\end{array}\right.
\\
\end{array}
\end{equation} 
This set of functions is commonly used in the theory of stability and 
buckling of elastic columns \cite{Beams}. 

The eigenvalues of the boundary-value 
problem (\ref{orts1}) satisfy the dispersion equation
\begin{equation}
\label{disp}
1-\cos(\lambda_i L)=\frac{\lambda_i L}{2}\sin(\lambda_i L)
\end{equation}
So, one can see that the set of functions consists of two subsets. The first subset is defined
by the eigenvalues given by
\begin{equation}
\label{disp1}
\begin{array}{c}
\cos(\lambda_i L)=1\\
\sin(\lambda_i L)=0.
\end{array}
\end{equation}
That is
$$
\lambda_i=\pi\cdot(i+1)/L,\,\,i=1,3,5\ldots .
$$ 
In this case the associated eigenfunctions are given by
\begin{equation}
\label{set1}
{\mathcal X}_i=\cos(\lambda_i z)-1.
\end{equation}

And the second subset is defined by
\begin{equation}
\label{disp2}
\begin{array}{c}
\cos(\lambda_i L)=\frac{4-(\lambda_i L)^2}{4+(\lambda_i L)^2}\\
\sin(\lambda_i L)=\frac{4\lambda_i L}{4+(\lambda_i L)^2}.
\end{array}
\end{equation}
The associated eigenfunctions are given by
\begin{equation}
\label{set2}
{\mathcal X}_i=\cos(\lambda_i z)-1+2z-\frac{2}{\lambda_i L}\sin(\lambda_i z).
\end{equation}
The first eigenvalue of this subset is equal to $\lambda_2=8.986819/L$. Further eigenvalues
are approximately given by $\lambda_i\simeq\pi\cdot(i+1)/L,\,\,i=4,6,8\ldots$.

Now, with the help of (\ref{ortf}) one can obtain from  (\ref{OneS}) a criterion 
when the instability first appears
\begin{equation}
\label{incr1}
\alpha (T-T_{0})ES-\frac{\mu _{0}}{4\pi} I^{2}(\ln (\lambda _{1}a)+
0.14)>EJ\lambda_1^2.  
\end{equation}
The first term on the left hand side of (\ref{incr1}) 
is responsible for the thermal expansion effect, while the second term represents 
magneto-elastic buckling. It should be noted that while we are considering both effects 
simultaneously, for parameters relevant to wire
explosion experiments these terms have different orders of magnitude.

The criterion obtained shows when the first eigenmode of (\ref{One}) and
the system as a whole
become unstable, $\lambda _{1}=2\pi/L$ being the corresponding eigenvalue.
In a similar manner, substituting other eigenvalues $\lambda_i$ for $\lambda_1$, 
one can obtain respective criteria for higher modes as well.

The set (\ref{orts1}) being very usefull in the linear stability analysis
of system (\ref{One}) doesn't seem to be very relevant to study the increments of the instability
since 
$$
\int_0^L {\mathcal X}_i{\mathcal X}_k dz\neq 0,\,\, i\neq k.
$$
But, it would be interesting to obtain an estimate of the increments of the instability. 
For this purpose, 
we will substitute a solution to (\ref{One}) in the form $X=A_i(t)X_i$, i.e assuming that only
one mode is present.  
Then, for the increment $\gamma_i$, $(A_i(t)\sim\exp(\gamma_i t))$, one gets
\begin{equation}
\gamma_i^2=\frac{E}{\rho}\frac{\lambda_i^2}{2}\frac{L}{\int_0^LX_i^2 dz}
\left\{\Xi-\lambda_i^2 \frac{J}{S}\right\}
\end{equation}
$$
\Xi=\alpha (T-T_0)-\frac{\mu_0 I^2}{4\pi ES}
\left\{0.14+\ln(a\lambda_i)  \right\}
$$
or since $\int_0^LX_i^2 dz\simeq L$
\begin{equation}
\label{incr2}
\gamma_i\simeq\sqrt{\frac{E}{\rho}}\lambda_i
\{\Xi-\lambda_i^2 \frac{J}{S}  \}^{1/2}
\end{equation}

As is seen from (\ref{incr2}), 
the increment has a maximum at $\lambda_i=\lambda_{ext}$ defined by
the equation 
\begin{equation}
\label{ext}
\lambda_{ext}^2=\frac{S}{J}\frac{\alpha (T-T_0)}{2}-
\frac{\mu_0 I^2}{8\pi EJ}\{0.64+\ln(a\lambda_{ext}) \}.
\end{equation}
As we will see further, for the range of parameters used 
in the wire explosion experiments, the contribution from the terms 
due to magnetic force can be neglected during the initial stage and within this approximation, 
using explicitly $J=\pi a^4/4$ and $S=\pi a^2$, 
\begin{equation}
\label{lextr}
\lambda_{ext}=\sqrt{\frac{2\alpha(T-T_0)}{a^2}}.
\end{equation}
And, since $\lambda_i\simeq \pi\cdot(i+1)/L$, 
\begin{equation}
\label{iextr}
i_{ext}=\sqrt{\frac{2\alpha(T-T_0)}{a^2}}\frac{L}{\pi}-1.
\end{equation} 
On the other hand, the maximal $\lambda_{lim}$ at which instability may exist is  
obtained from (\ref{incr2}) to give 
\begin{equation}
\label{maxlam}
\lambda_{lim}\simeq 2\sqrt{\frac{\alpha(T-T_0)}{a^2}}
\end{equation} 
with the same accuracy.
From (\ref{maxlam}), since $\lambda_i\simeq \pi\cdot(i+1)/L$, 
one gets for the maximal mode number
\begin{equation}
\label{maxi}
i_{lim}\simeq 2\sqrt{\frac{\alpha(T-T_0)}{a^2}}\frac{L}{\pi}-1
\end{equation}  
Thus the dominant mode lies
somewhere between the first unstable mode and the last one with the
maximal increment
\begin{equation}
\label{incrmax}
\gamma_{ext}=\sqrt{\frac{E}{\rho}}\frac{\alpha(T-T_0)}{a}.
\end{equation}

From the expression (\ref{iextr}) one can see that
higher modes are likely to become dominant 
in the instability spectrum with increase in the wire length. 
As a result, the wire shape can take a rather intricate form. 
It seems that the instability on higher modes has been
observed experimentally by Graneau, \cite{Graneau87}. 
At the experimental conditions he used, i.e. aluminium wire, 
$L=1\,\mbox{m}$, $a=0.6\,\mbox{mm}$, from (\ref{lextr}) one gets 
$\lambda_{ext}\simeq 294$ or the corresponding mode number $i_{ext}=93$
at $T=T_{melt}$. 
The photographs of
the wire shape after the current had been switched 
off showed clearly the appearance of at least $5$ bulges 
on a small part of the wire.

We will refer further to the sets of parameters presented in Table I 
corresponding to different wire lengths and different currents. 
One should note that two sets, [B] and [E], are relevant for
wires and currents used in the experiments \cite{Graneau87} and 
\cite{Nasilov64} respectively. 

\vspace*{0.4cm} \begin{center} Table.I \end{center}

\noindent
\begin{tabular}{|p{0.2in}|p{0.9in}|p{0.6in}|p{0.6in}|p{0.7in}|p{0.7in}|p{0.7in}|p{0.7in}|}
\hline
&material& $a$&$L$&$I$&$P_1$&$P_2$&$P_3$\\
\hline
A&aluminium& 0.6 mm& 0.05 m& 5 kA& $1.19\times 10^3$&6.11 &114 \\
\hline
B&aluminium& 0.6 mm& 1 m& 5 kA&$1.19\times 10^3$ &13.6 &0.28 \\
\hline
C&aluminium& 0.6 mm& 1 m& 2 kA& $1.19\times 10^3$&2.18 &0.28 \\
\hline
D&aluminium& 0.6 mm& 0.3 m& 8 kA& $1.19\times 10^3$&34.8 &0.28 \\
\hline
E&copper& 0.5 mm& 1 m&500 A &$1.84\times 10^3$ &0.14 &0.24 \\
\hline
\end{tabular}
\vspace*{0.5cm}

For all the represented sets, 
the criterion (\ref{incr1}) is fulfilled well enough. Indeed, we have put in the Table I 
the corresponding values of the terms in (\ref{incr1}), designating them as
$P_1$, $P_2$ and $P_3$ from left to right. The estimations have been
done at $T=T_{melt}$, $T_{melt}$ being the melting temperature;
$T_{melt}=660^{\circ}\, \mbox{C}$ for aluminium and 
$T_{melt}=1085^{\circ}\, \mbox{C}$ for copper.
The characteristic increment (\ref{incrmax}) at the same conditions, in
the case [B] for instance, is $\gamma_{max}\simeq 10^5\, \mbox{sec}^{-1}$,
which is much greater than the temperature increment 
$\gamma_{T}\simeq 7\times 10^2 \,\mbox{sec}^{-1}$. Thus, even though the estimation has 
been done 
in the linear approximation, it is clear that the instability has sufficient time to develop
before the wire reaches the melting point. 

In all the cases, as is
seen, the major contribution is from the thermal expansion effect, while the 
influence of the magnetic-force terms can be neglected 
during the initial stage 
(this, of course, can be estimated directly from (\ref{One}) as well).
Two terms, $P_1$ and $P_2$, for instance in case [B],
become equal only at $T-T_0=6.6^{\circ} \,\mbox{C}$. 
The time it takes for the temperature to be driven through this range is 
$\Delta t\simeq 30\,\mu\mbox{sec}$. The characteristic time 
for the instability to develop for this
temperature difference is much longer,
$\gamma_{max}^{-1}=800\, \mu\mbox{sec}$.

\subsection{Stability of an initially bent wire}
Consider a wire which has an initial form $X_0=\sum_i A^0_{i}{\mathcal X}_i$,
where the functions $\{{\mathcal X}_i\}$ is the set (\ref{orts1}) and $A^0_i$ are given weight 
coefficients. One needs to stress,
that with the accuracy the system (\ref{One}) was derived, $|X_0|>a$ must always
be the case. 
Performing analysis similar to that leading to equation
(\ref{incr1}) and neglecting the magnetic force for the sake of 
simplicity, for a mode $k$ one gets
\begin{equation}
\label{bentwire}
\lambda_k^2 (A_k-A_{k}^0)EJ-A_k ES(\alpha \Delta T)=0.
\end{equation}
From (\ref{bentwire}) one can see that there is always a stationary state 
which differs from $X_0$
at any nonzero value of the parameter
$\alpha \Delta T$. Thus,
those modes which have 
$A^0_{k}\neq 0$ are always unstable. The result is not 
surprising, since from the physical point of view it is obvious that
a constantly heated wire will change its form owing to expansion.
On the other hand, those modes which have $A^0_{k}=0$ become unstable
at some value of the parameter $\alpha\Delta T$, so that buckling instability
is still possible on these modes. 

\section{Qualitative stress analysis.}

Once the instability occurs, all the potential energy comprised into the
compressed wire can be quickly released. The characteristic value of the
longitudinal stress $\tau _{zz}$ which may be accumulated during the
preconditioning is within the interval between the maximum value $\tau
_{zz}^{max}=\alpha (T_{melt}-T_{0})E\sim 10^{3}\,\mbox{MPa}$ and 
the minimum value $\tau _{zz}^{min}=E
\frac{J}{S}\lambda _{0}^2 \simeq 0.3\div 30\,\mbox{MPa}$ 
corresponding to the onset of the
instability. The estimation has been done for an aluminium wire of
radius $a=0.6\,\mbox{mm}$ and the length spanning the range 
$1\,\mbox{m}>L>0.05\,\mbox{m}$. 

One can see, that the maximum value itself is very high, about
10 times higher than the ultimate strength value. On the other hand, the
minimum value can be quite small depending on the wire length. 
The average
transverse stress is given by $<\tau _{xx}>=
\frac{G}{S}X^{'}-E\frac{J}{S}X^{'''}$. 
For small deflections, $|X|/\xi_L\ll 1$, it is in general smaller 
then the longitudinal one. 

From the qualitative point of view it is evident that the maximal stress
energy can be accumulated if the temperature rises sufficiently quickly.  
The temperature rise time must be shorter or comparable with the
rise time of the instability. The most dramatic result might be
expected if the wire is heated up to the temperature just slightly 
below the melting point. Then the current is switched off thus allowing the
instability to develop without melting the material. On the other hand,
if the current is low and, as a consequence, the temperature rise time
is low too, then all the accumulated energy can be released at the onset of
the instability by the first mode which is becoming first unstable.
This case corresponds to the lower limit of the estimated stress 
value $\tau_{zz}^{min}$. 

Thus, already now, from the linear analysis, it is
obvious that many scenarios of the instability are possible. 
Dynamically, many modes can be excited simultaneously
while the temperature $T$ increases from $T_{0}$ up to the melting point 
$T_{melt}$. Even though the first mode becomes first unstable it might
happen that further other modes play dominant role creating rather complex dynamical 
behaviour.  Any particular
pattern, of course, depends on temporal characteristics such as the
ratio between instability increment and the temperature rise time.

\section{\protect\bigskip Numerical results}

Now, there are several further questions. 
What character will the instability have on the nonlinear stage? 
How high longitudinal
and transverse stresses could be obtained in this situation? 
To answer these questions we 
solve the nonlinear equation (\ref{One}) numerically. 

\subsection{Numerical algorithm}
The numerial method of the solution of equation (\ref{One}) with the
boundary conditions (\ref{clends}) is based on the complete set of orthogonal
functions (\ref{orts1}). 

By expanding any solution to (\ref{One}) into the series of the functions $\{{\mathcal X}_i\}$, 
i.e $X=\sum_i A_i {\mathcal X}_i$, the partial differential equation turns into a system of
ODEs with respect to time. They are  
\begin{equation}
\label{numeric}
\sum_jB_{ij}\frac{d^2 A_j}{dt^2}+A_i\frac{\lambda_i^4}{2} + A_i\frac{\lambda_i^2}{2}
\left\{\beta_a
(\tilde{l}-\tilde{l_0}-\alpha[T(t)-T_0])+\beta_{B}(0.14+\ln(\lambda_i a/L))\right\}=0
\end{equation}
where the coefficients are $\beta_a=4L^2/a^2$, 
$\beta_{B}=\frac{\mu_0 I^2 L^2}{4\pi EJ}$. The symmetric matrix $B_{ij}$ is given by 
$B_{ij}=\int_0^1 X_i X_j dz$. The length increment is  
$\tilde{l}-\tilde{l_0}=\frac{1}{4}\sum_j(A_j^2-{A^0_j}^2)\lambda_j^2$.
The functions
$X$, $X_0$ and the variable  $z$
have been normalised on $L$ and $t$ has been normalised by 
$t_0=\sqrt{\frac{\rho S}{EJ}}L^2$.
In the numerical simulations, the infinite series have been trancated at the maximal
unstable mode defined by (\ref{maxi}).

\subsection{Flexural vibrations of initially straight wires}
In the case of initially straight wires the initial conditions are $A_i(0)=0$.
To excite the instability  one needs to seed some initial noise level. To simulate
noise present an additional fluctuation force in the equation (\ref{One}),
$\delta F=\delta F_0\delta f(t)\sum_i \exp(-\xi_i^2/\xi_1^2){\mathcal X}_i$, has been added,
with $\delta F_0=4\,\times 10^{-4} \,\mbox{newton}$. The function $-1<\delta f(t)<1$ has been
calculated by means of a random number generator at each step of calculations over 
time. This force can give rise to deflection of an aluminium wire with 
$L=1\,\mbox{m}$ and $a=0.6\,\mbox{mm}$ from a straight line by approximately $X\sim 0.1\, a$.

The current profile in the simulations has been taken in the form\linebreak[4]
$I(t)=I_0\ sinh(t/t_0)/cosh(t/t_0)$, where $t_0$ is the current rise time.
In all runs $t_0=30\,\mu\mbox{sec}$. This time
is about twice as high as the skin time for aluminium wires with
$a=0.6\,\mbox{mm}$, $t_{skin}=a^2\sigma\mu_0\simeq 17 \,\mu\mbox{sec}$. 
We also kept $T_0=300^{\circ}\,\mbox{K}$ and 
calculations stopped once the temperature had reached the melting point
$T=T_{melt}$. We designated this moment by $t=t_{e}$.

We have performed simulations for different conditions: different wire
lengths, different currents and different wire materials.  
All the sets used are presented in Table I. 

As the first example, let's consider a short
aluminium wire carrying $5 \,\mbox{kA}$ current, case [A]. For these
parameters from the linear analysis one would expect a few
unstable modes to develop, $i_{lim}=6$. The results obtained are presented in Fig.1. 
We have plotted in the first two frames 
the deflection $X$ as a function of $z$ 
and the spectrum $<A(t)_{i}^2>$ both taken at $t=t_e$.
In the last two frames, we have plotted the longitudinal $\tau_{zz}$ 
and transverse $<\tau_{xx}>$ stress components as functions of time.
The spectrum has been calculated by means of averaging 
over time around $t=t_e$, i.e.
$<A(t_e)_i^2>=\int_{t_e-\Delta}^{t_e}A(t)_i^2 dt$, where $\Delta$ is
chosen to be greater than the period of nonlinear vibrations.
Both the spectrum, Fig.1b and
the wire shape, in the form being close to a simple arc, Fig.1a, 
show that the first mode dominates during the nonlinear stage 
of the instability. 
Temporal behaviour of the longitudinal and transverse
stress components demonstrates developed nonlinear 
flexural vibrations, Fig.1c and Fig.1d. The appearance
of the instability is clearly seen at $t\simeq\,900\mu\mbox{sec}$ in both figures.
The observed longitudinal stress has both compressive and tensile components.
The instability appeared after strong compression, the maximal compressive stress
being $429\,\mbox{MPa}$. 
The maximal tensile longitudinal stress is amounting to $\tau_{zz}\simeq
244\,\mbox{MPa}$. This value is well above the ultimate stress value for
aluminium. The observed transverse stress, as has been expected, is 
lower then the longitudinal one, $<\tau_{xx}>\simeq 55\,\mbox{MPa}$.

Let's consider now a longer wire, case [B]. This case is relevant to the conditions of the
Graneau's experiments \cite{Graneau87}. One might expect, 
in accordence with our
qualitative analysis, more active modes in
the instability spectrum to develop with respect to the previous case, $i_{lim}=129$. 
Indeed, from Fig.2a and Fig.2b one can observe that the spectrum becomes 
very rich with the major contribution from mode $n=27$ at the melting point. 
The tensile longitudinal and transverse stresses are amounting to 
$\tau_{zz}=122\,\mbox{MPa}$ and $<\tau_{xx}>=92\,\mbox{MPa}$ in this case, see
Fig.2c and Fig.2d. Both of them are well above the ultimate strength of aluminium.
Thus, one might expect the first wire break to appear just after buckling occured,
at $t=1000\,\mu\mbox{sec}$. 

Let's investigate now the effect of variations 
of current. In case [C], which is a replica of case 
[B] but with lower current, two modes $n=5$ and $n=7$ become dominant, Fig.3a. 
This fact indicates, if we compare two spectra in cases [B] and [C],
that the main channel of the accumulated energy to release 
is going now through the modes with lower numbers. Thus, one would anticipate
a lower accumulated stress according to our qualitative stress analysis. 
Indeed, we have observed that both tensile
longitudinal and transverse stresses 
reduced to $\tau_{zz}=70\,\mbox{MPa}$ and $<\tau_{xx}>=23\,\mbox{MPa}$, 
Fig.3b and Fig.3c.

On the other hand, an increase of the current leads to
higher resultant tensile stress value. In case [D] which is 
similar to set [B] but with a higher current, the 
longitudinal tensile stress is $\tau_{zz}\simeq 233\,\mbox{MPa}$ in maximum, Fig.4.
It should be noted that because of faster 
heating the preliminary compression before the instability
developed is much more stronger, namely $851\,\mbox{MPa}$, and the instability
occurred just before the melting point.
It appears that at higher current the wire will be molten before the buckling occurs.

Let's consider now the Nasilovski's
experiment with a long copper wire, case [E]. 
This experiment has been carried out at quite a
low current $I=500 \,\mbox{A}$ in comparison with the all previously 
considered cases. As a result, one might expect low stress
values compared to previous cases. Also, a few first modes
must be dominant. Indeed, that is just the case, see Fig.5a.
The observed values of stress indeed become quite low. The longitudinal stress is reaching 
$\tau_{zz}\simeq 27\,\mbox{MPa}$ in maximum, Fig.5b, while the transverse stress is just
about $4\,\mbox{MPa}$. It means that the wire can be broken only if it
was heated up enough.

\subsection{Flexural vibrations of initially bent wires}
In the above, we considered somewhat idealised situation, when the undeformed wire
had the shape of a straight rod. In reality, the wire
might have had initially any form. For instance, in the case of horizontal
positioning of the wire it might be bent by gravitational force. Once the
initial dislocation becomes grater than the radius, the character of the
instability changes as has been discussed. Indeed, if $X_0\neq 0$ 
the initial state is unstable from the beginning.
But, if the temperature risetime is
shorter than the period of flexural vibrations then those modes which contribute into
$X_0=\sum_iA^0_i{\mathcal X}_i$ can be excited directly. 
Moreover, buckling on the other modes is still 
possible. To estimate the stresses developed  
simulations for case [B] have been performed but 
with initial conditions given by $X_0=A^0_1 {\mathcal X}_1$ with $A^0_1=20 a$.
That is, the wire was initially shaped like the first mode of the set 
({\ref{orts1}). The choice of the amplitude $A^0_1$ follows from 
estimations for a horizontally positioned wire of this
length. The wire in this case has a form $X_0=\frac{qz^2(z-L)^2}{24EJ}$,
see the problem in \cite{LL} page 93, $q$ being the wire weight per unit length. 
Thus, the maximal displacement is expected to be ${X_{0}}_{max}\sim 18 a$.

The results of simulations are shown in Fig.6. 
As before, one can see developed nonlinear 
vibrations. The regular character of the vibrations points out to the fact that 
only one single mode $i=1$ is active in this case. This means that buckling on
the other mode didn't occur. 
The amplitude of tensile longitudinal stress is amounting for this particular
case to $\tau_{zz}\simeq 50\,\mbox{MPa}$. So, a sufficiently heated wire 
can be broken in this case as well.

\section*{Conclusions}

When an electric current passes through a thin metal wire with clamped 
ends, flexural elastic stress waves are induced owing to the Joule heating 
and the electromagnetic force. The Joule heating leads to thermal expansion 
of the wire material, which is the dominant mechanism of the excitation of 
vibrations. Under realistic experimental conditions the electromagnetic 
force is of minor importance.

Flexural vibrations in an initially straight wire may be excited as a 
result of the buckling instability. The energy accumulated in the wire 
during the initial stage in the form of a compressive stress is suddenly 
released. As a result, high tensile stress appears, which is sufficiently 
high to cause the fracture of the wire within 1 ms for all the cases 
considered.  The number of modes of the instability that are excited 
depends on the current magnitude and wire length. Depending on these 
parameters it is possible to excite just a single mode, which has a form of 
an arch, or actually any number of modes.

If a wire is slightly curved, which is a more realistic case than that of 
the straight wire, the buckling instability is still possible. The 
amplitude of modes, that are initially present in the curved wire, grows. 
The other modes can still be excited in a rapidly heated wire owing to 
buckling instability. The magnitude of tensile stress induced in a curved 
wire is clearly lower, but is still sufficient to induce a wire fracture on 
a millisecond timescale.

Three-dimensional effects, which may be induced by the external circuit, 
suspensions at the clamped ends, imperfections of the wire cross-section or 
a wire material, will increase the magnitude of tensile stress. These 
effects will lead to the coupling between flexural, longitudinal, and 
torsional modes. This study is beyond the scope of the present paper.

The experimental evidence, in general, is supportive of the mechanism 
described above. However, no direct comparison is possible, since previous 
experiments were of exploratory nature.

The model developed favours the view that the phenomenon of the 
wire fragmentation in the solid state can be explained without 
resorting to controversial Amp\`{e}re-Neumann electrodynamics. 

\section*{Acknowledgements}
This work has been supported by Engineering and Physical Sciences Research Council 
grant No.~GR/M07403.
The authors are grateful to Prof.~J.~Allen and Dr.~D.~Wall for usefull discussions.
A. Lukyanov would like to express his gratitude to O.V. Umnova for support.

\newpage
\section*{Captions}

Fig.1a The wire displasement $X/L$ as a function of $z/L$, case [A]. 

Fig.1b The spectrum of the wire vibrations at the melting point $<A_i^2>$ as a function of 
mode number $i$, case [A].

Fig.1c The longitudinal stress $\tau_{zz}$ as a function of time, case [A]. 

Fig.1d The transverse stress $<\tau_{xx}>$ as a function of time at $z/L=0.9$ at $z/L=0.9$, 
case [A]. 

Fig.2a The wire displasement $X/L$ as a function of $z/L$, case [B]. 

Fig.2b The spectrum of the wire vibrations at the melting point $<A_i^2>$ as a function of 
mode number $i$, case [B].

Fig.2c The longitudinal stress $\tau_{zz}$ as a function of time, case [B]. 

Fig.2d The transverse stress $<\tau_{xx}>$ as a function of time at $z/L=0.9$, case [B]. 

Fig.3a The spectrum of the wire vibrations at the melting point $<A_i^2>$ as a function of 
mode number $i$, case [C].

Fig.3b The longitudinal stress $\tau_{zz}$ as a function of time, case [C]. 

Fig.3c The transverse stress $<\tau_{xx}>$ as a function of time at $z/L=0.9$, case [C]. 

Fig.4 The longitudinal stress $\tau_{zz}$ as a function of time, case [D]. 

Fig.5a The spectrum of the wire vibrations at the melting point $<A_i^2>$ as a function of 
mode number $i$, case [E].

Fig.5b The longitudinal stress $\tau_{zz}$ as a function of time, case [E]. 

Fig.6 Flexural vibrations of an initially bent wire, longitudinal stress as a function of time.
$X_0=A_1^0{\mathcal X}_1$, $A_1^0=20\,a$. The wire and current parameters are 
relevant for case [B].
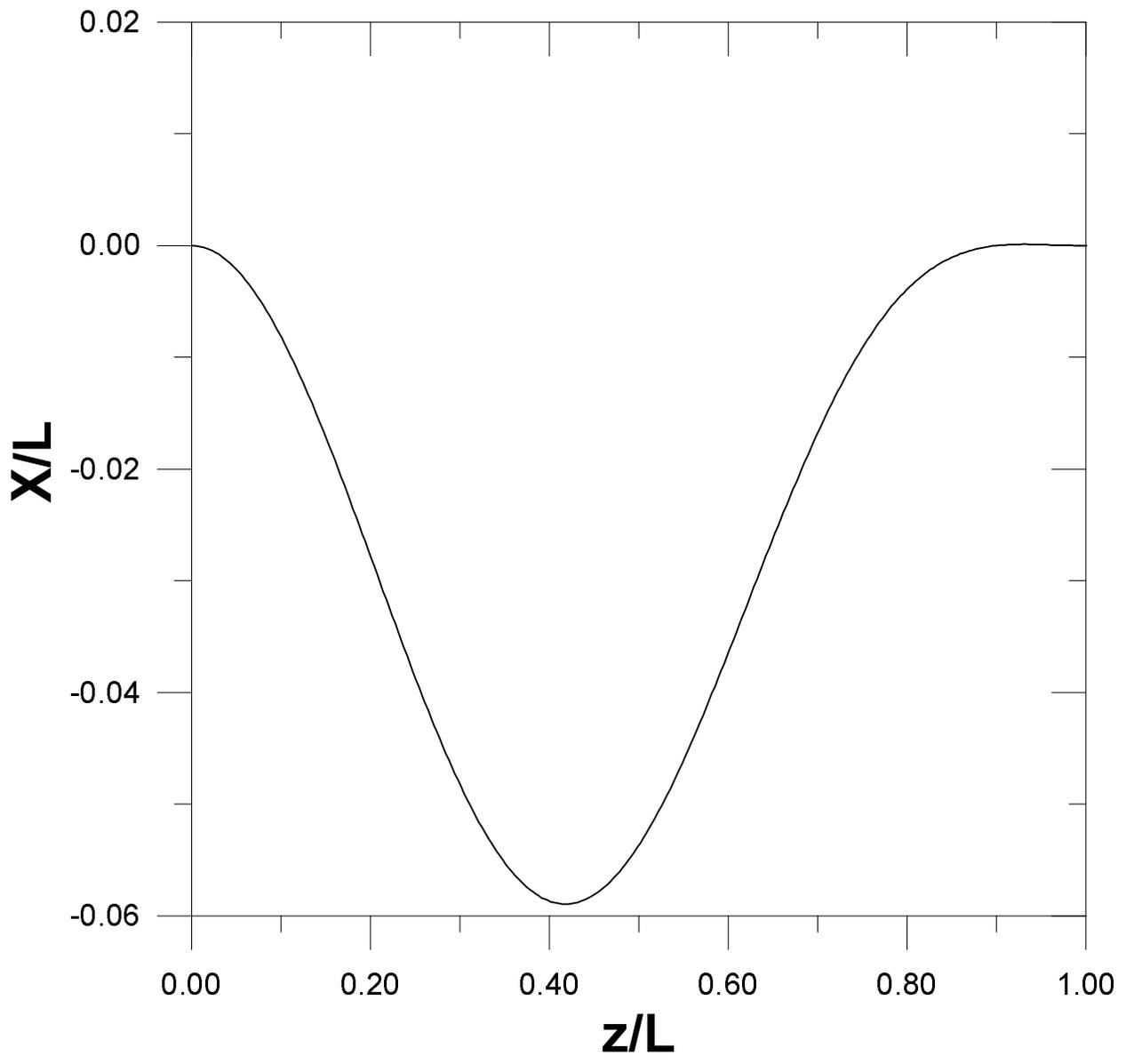
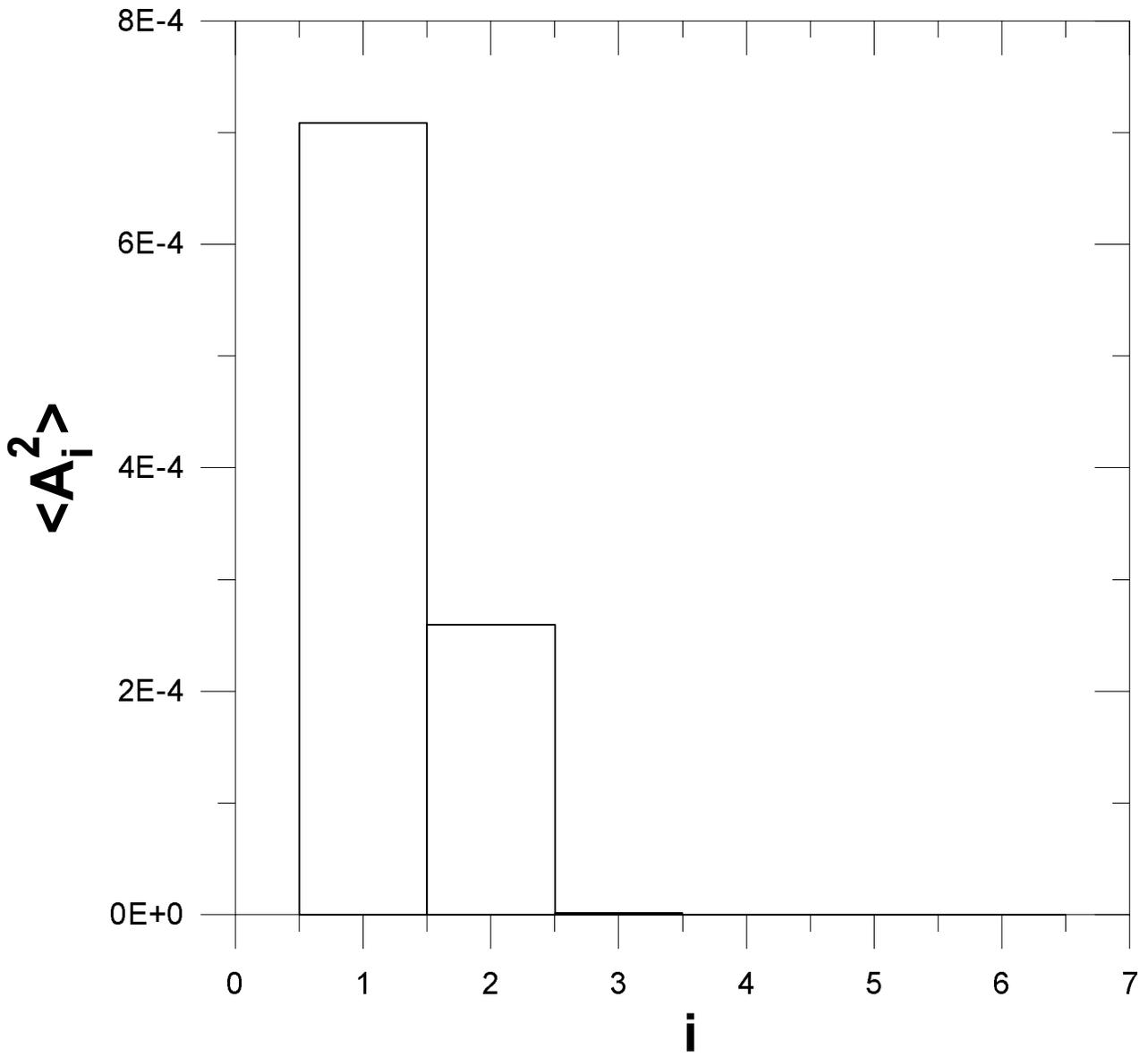
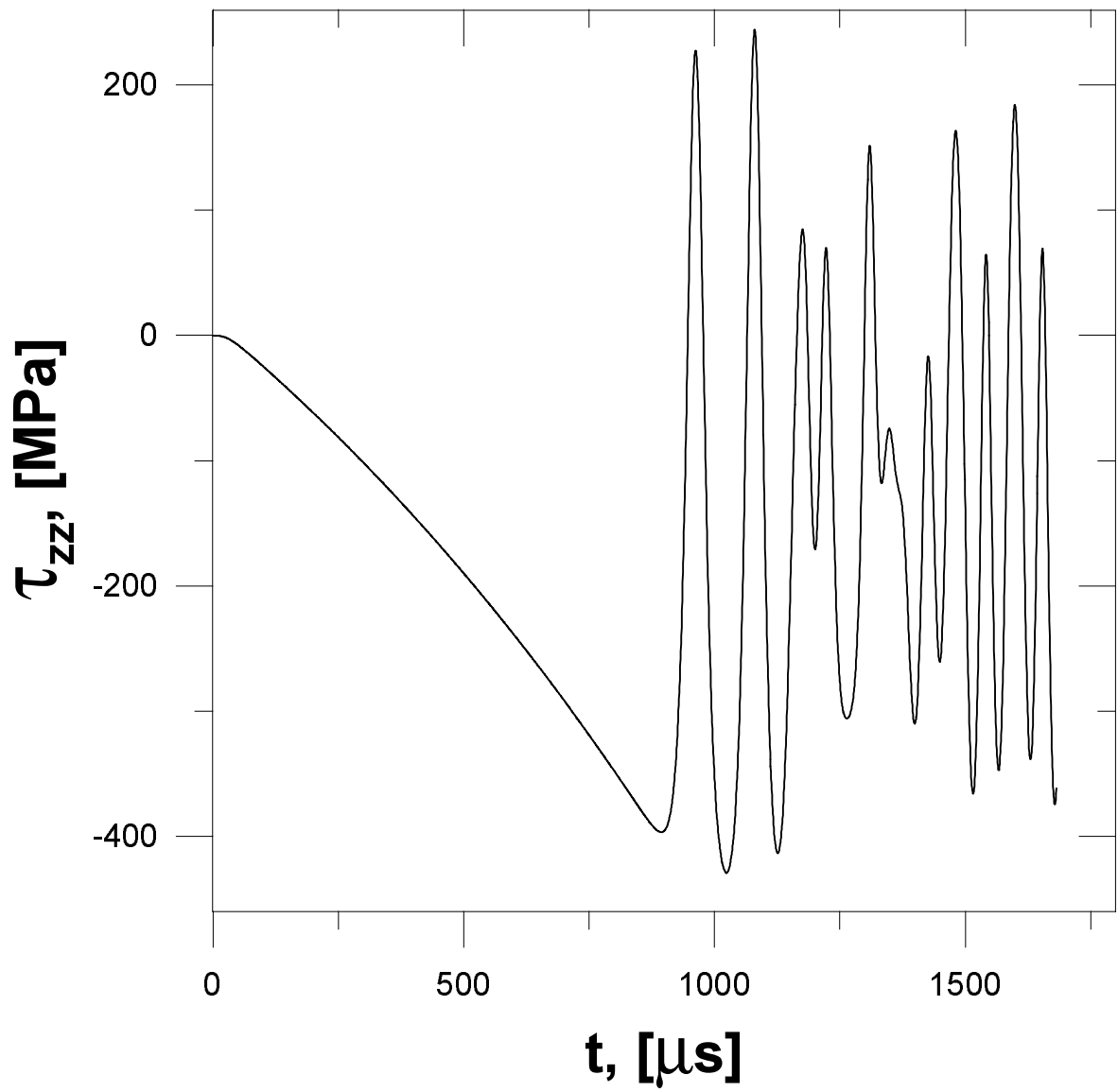
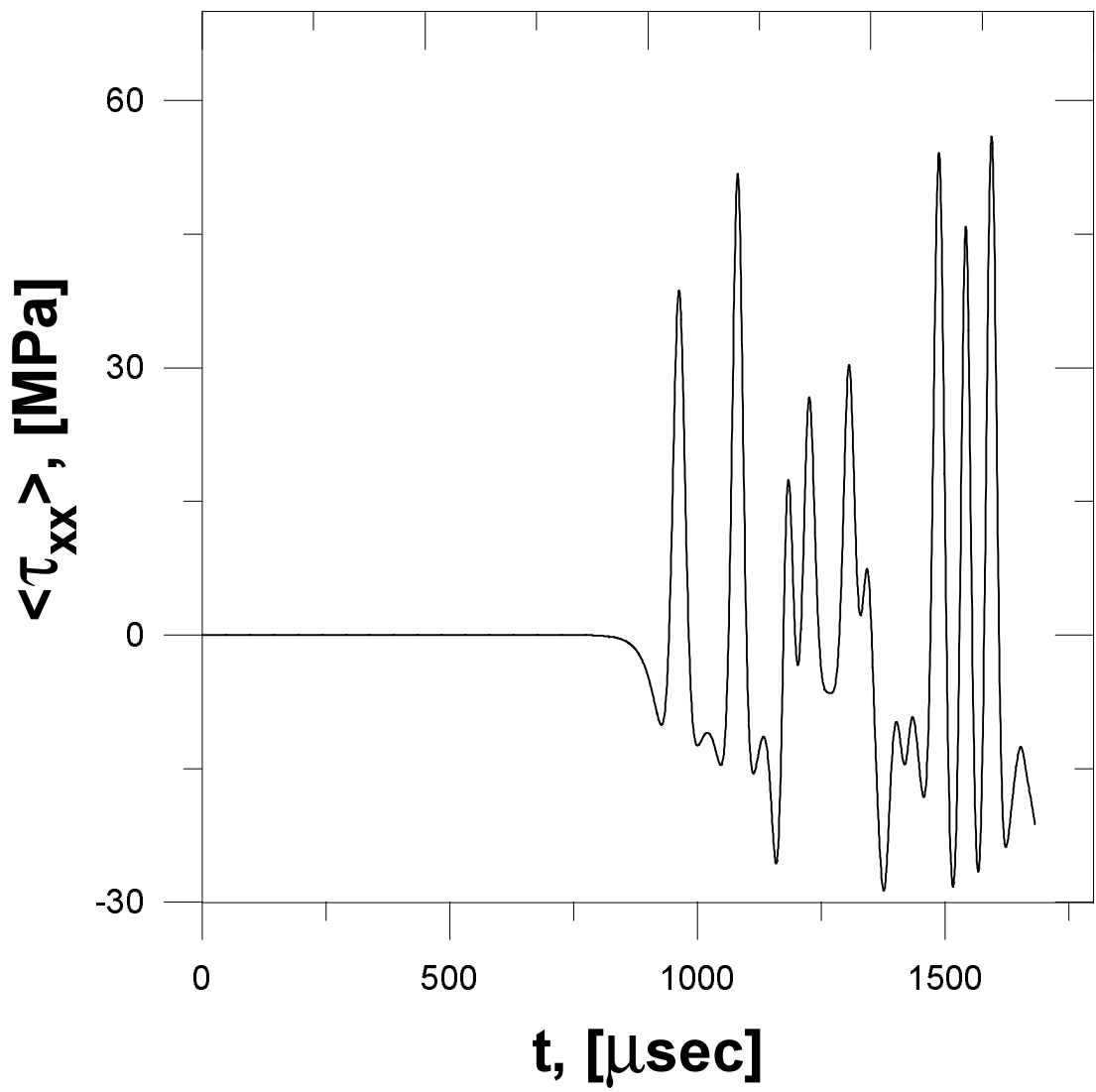
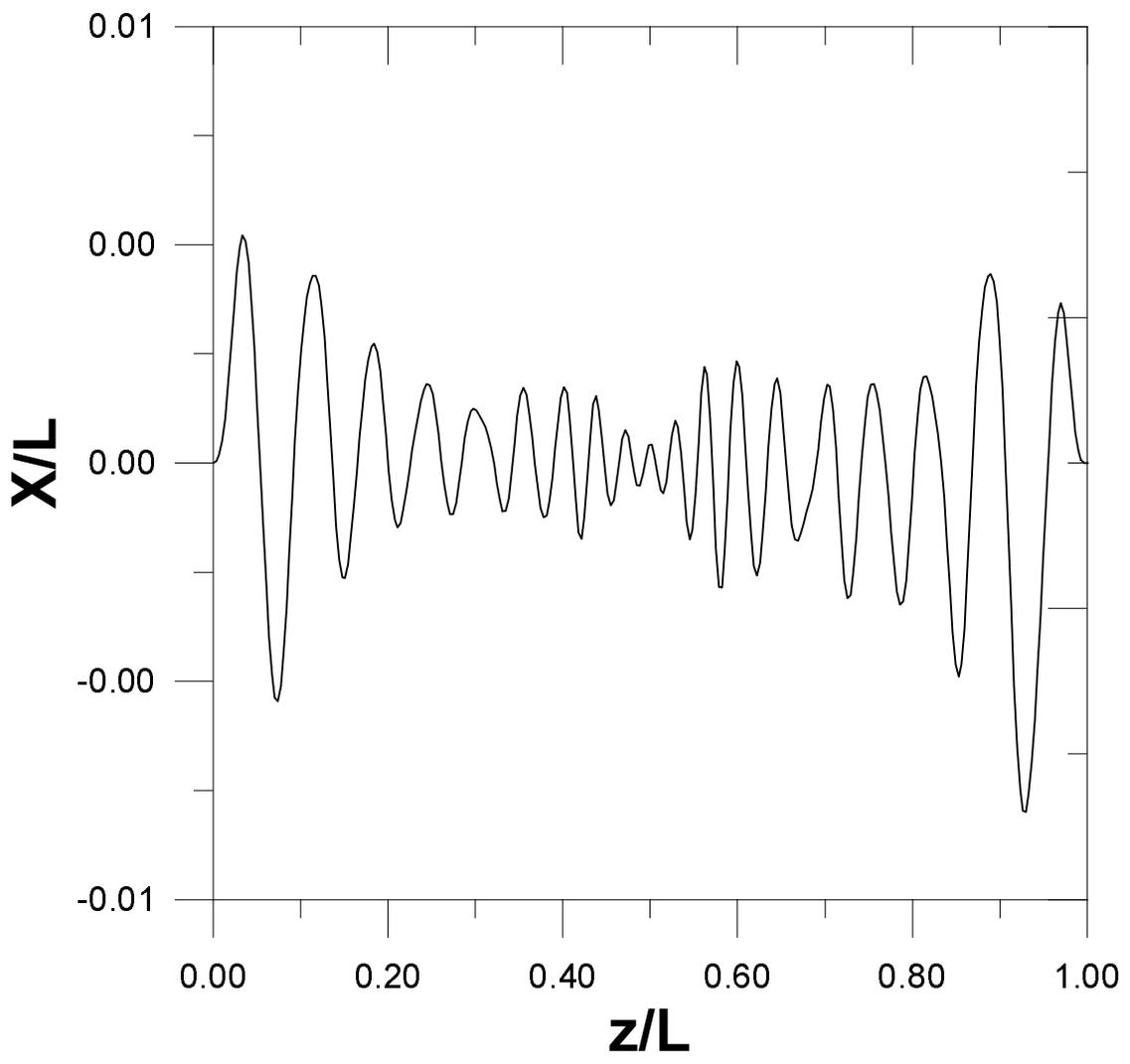
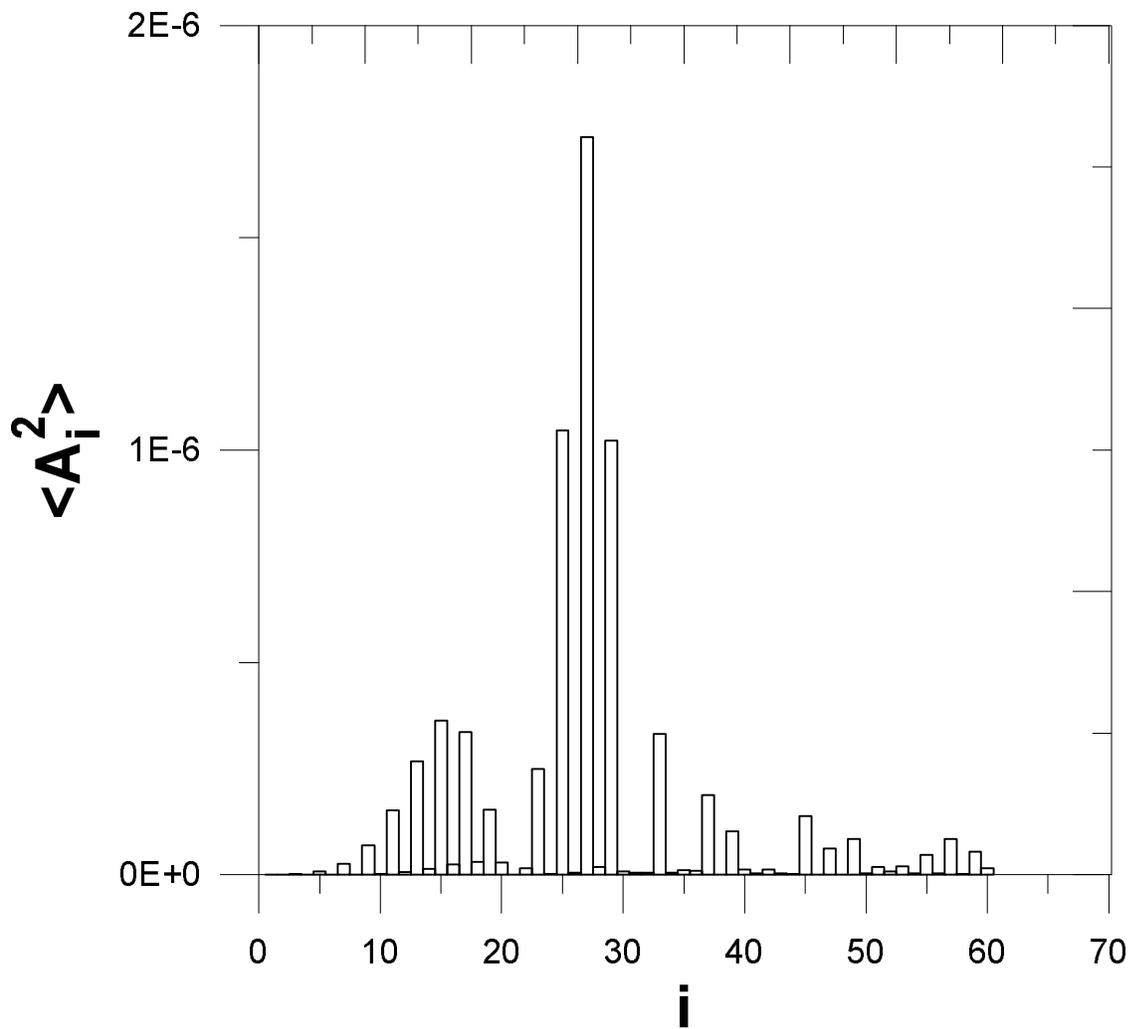
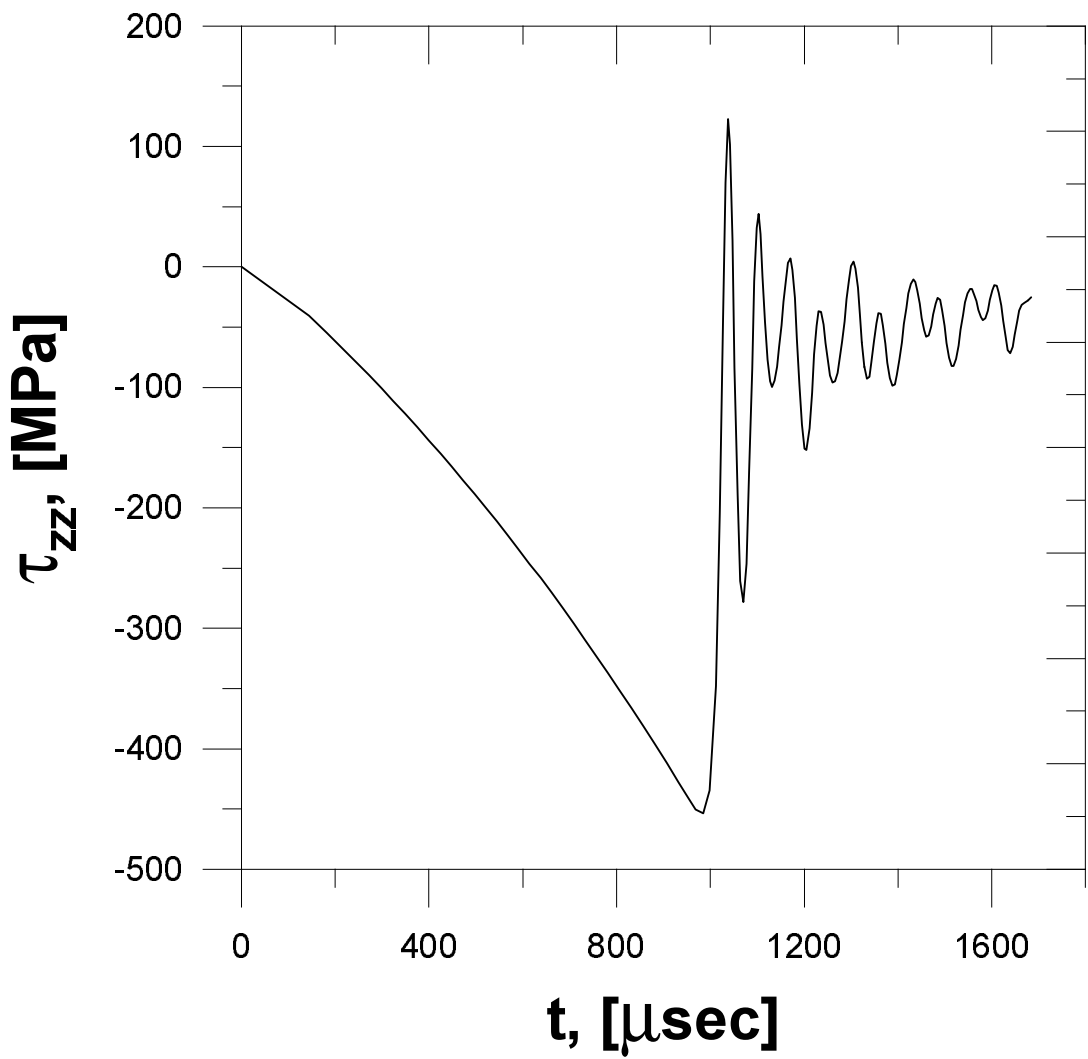
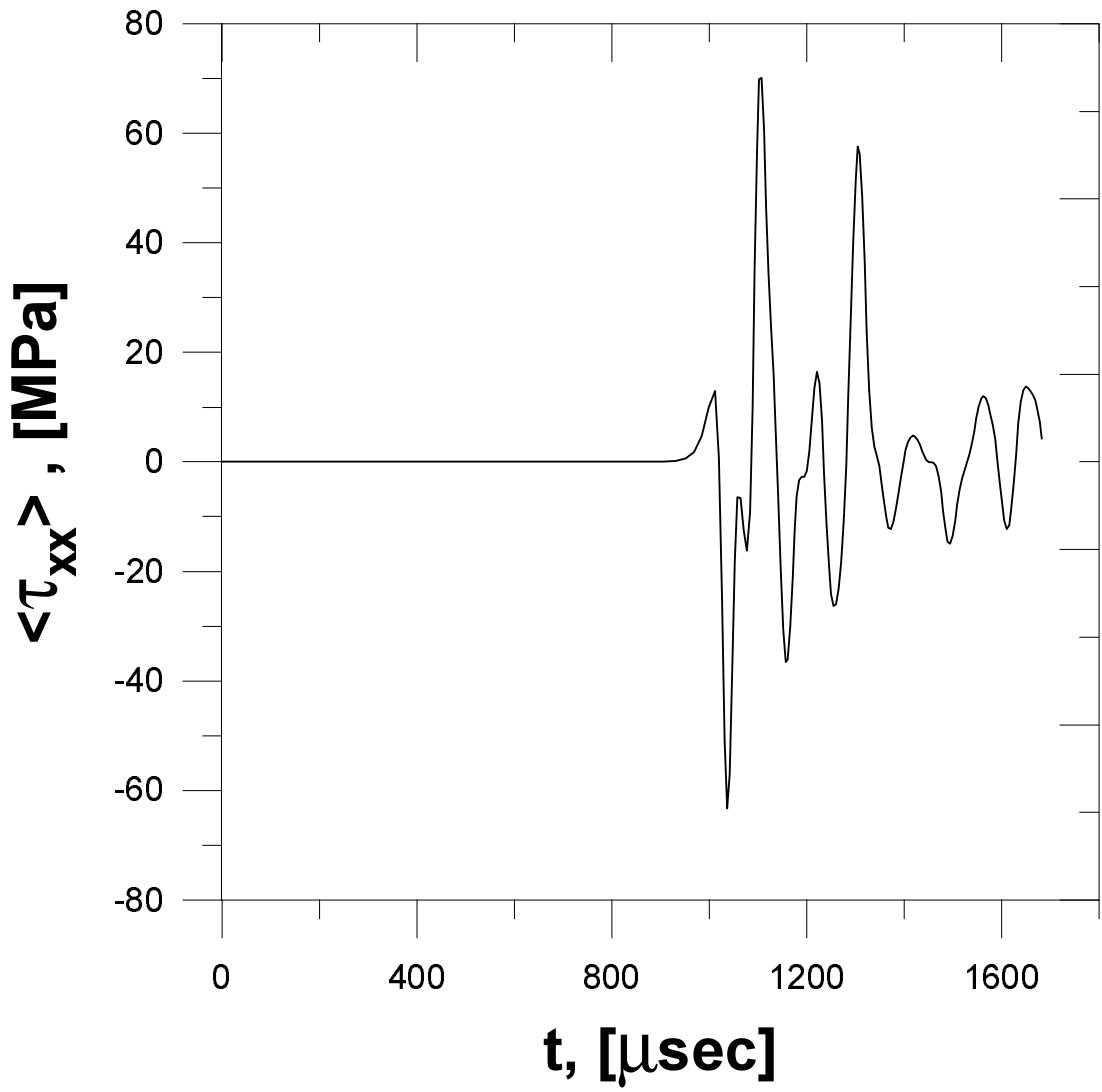
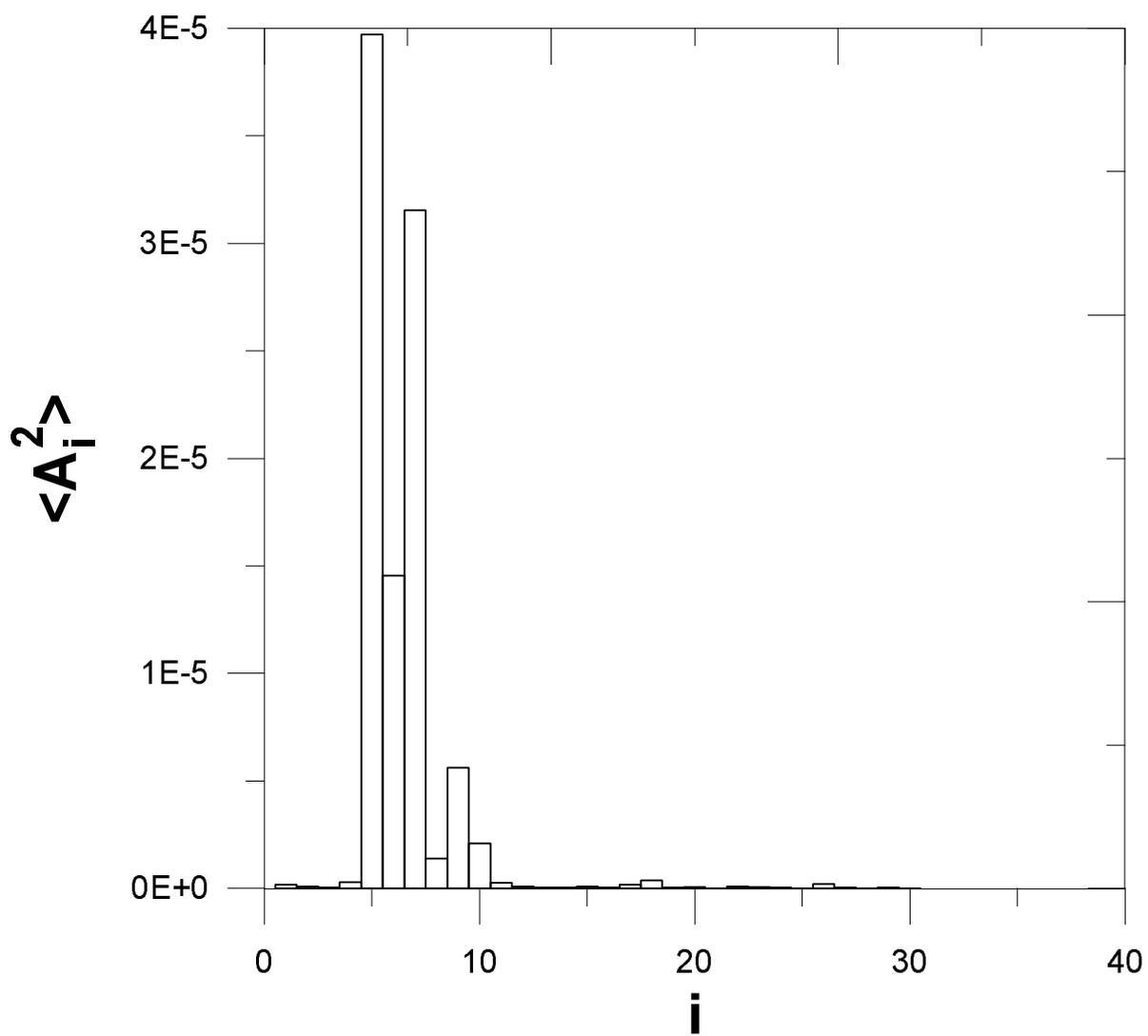
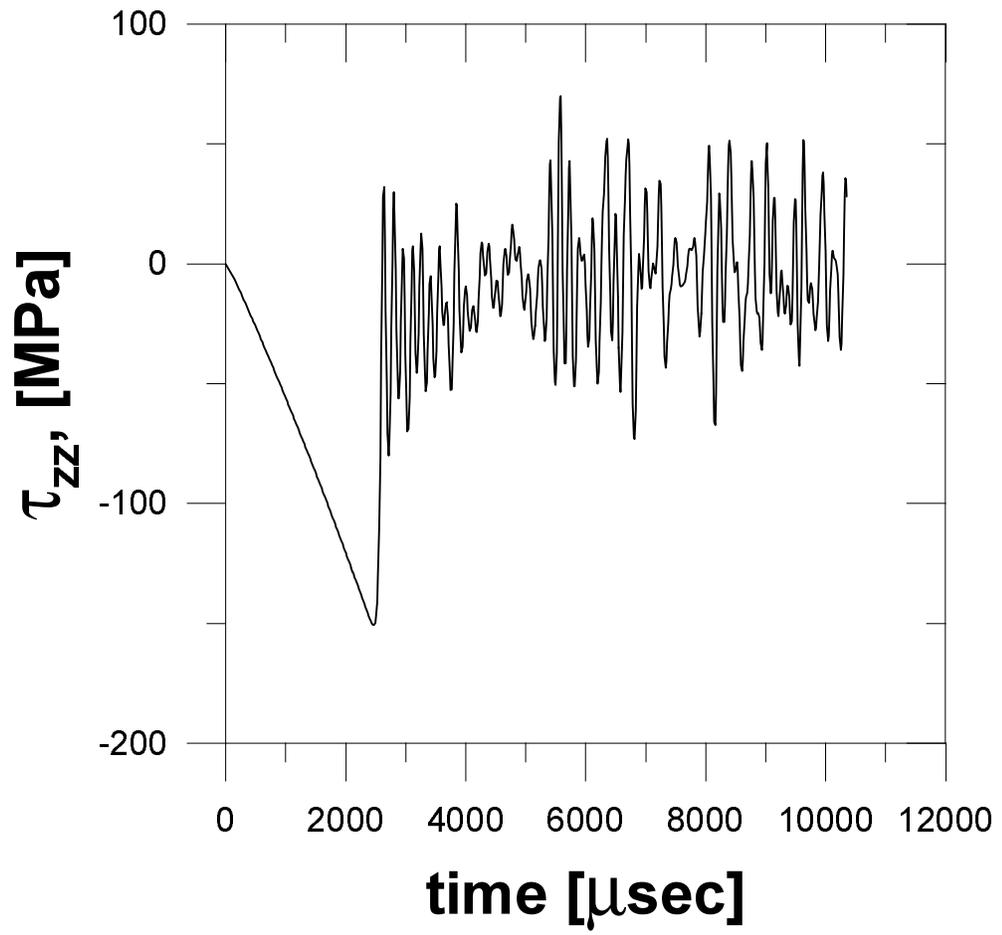
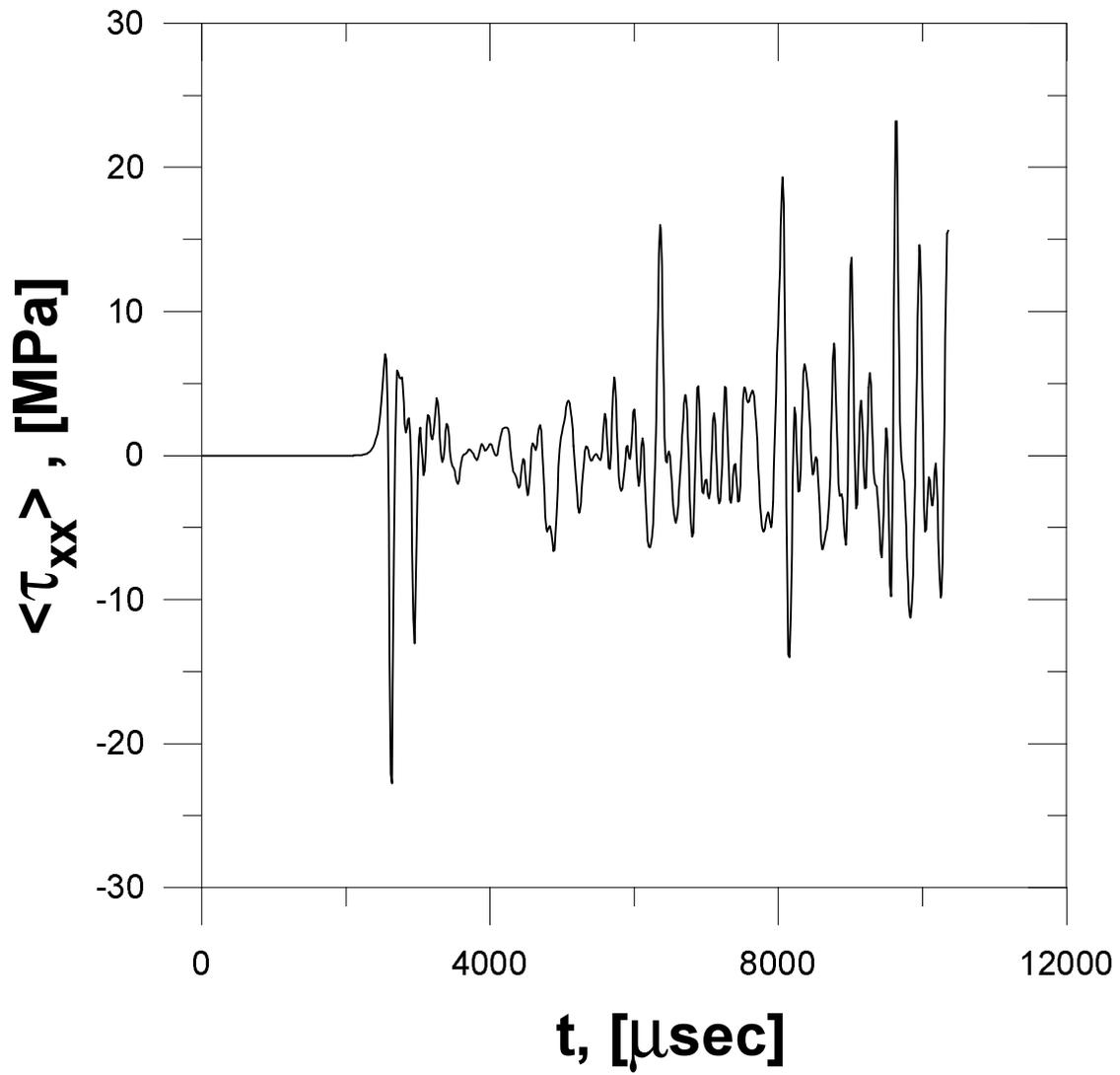
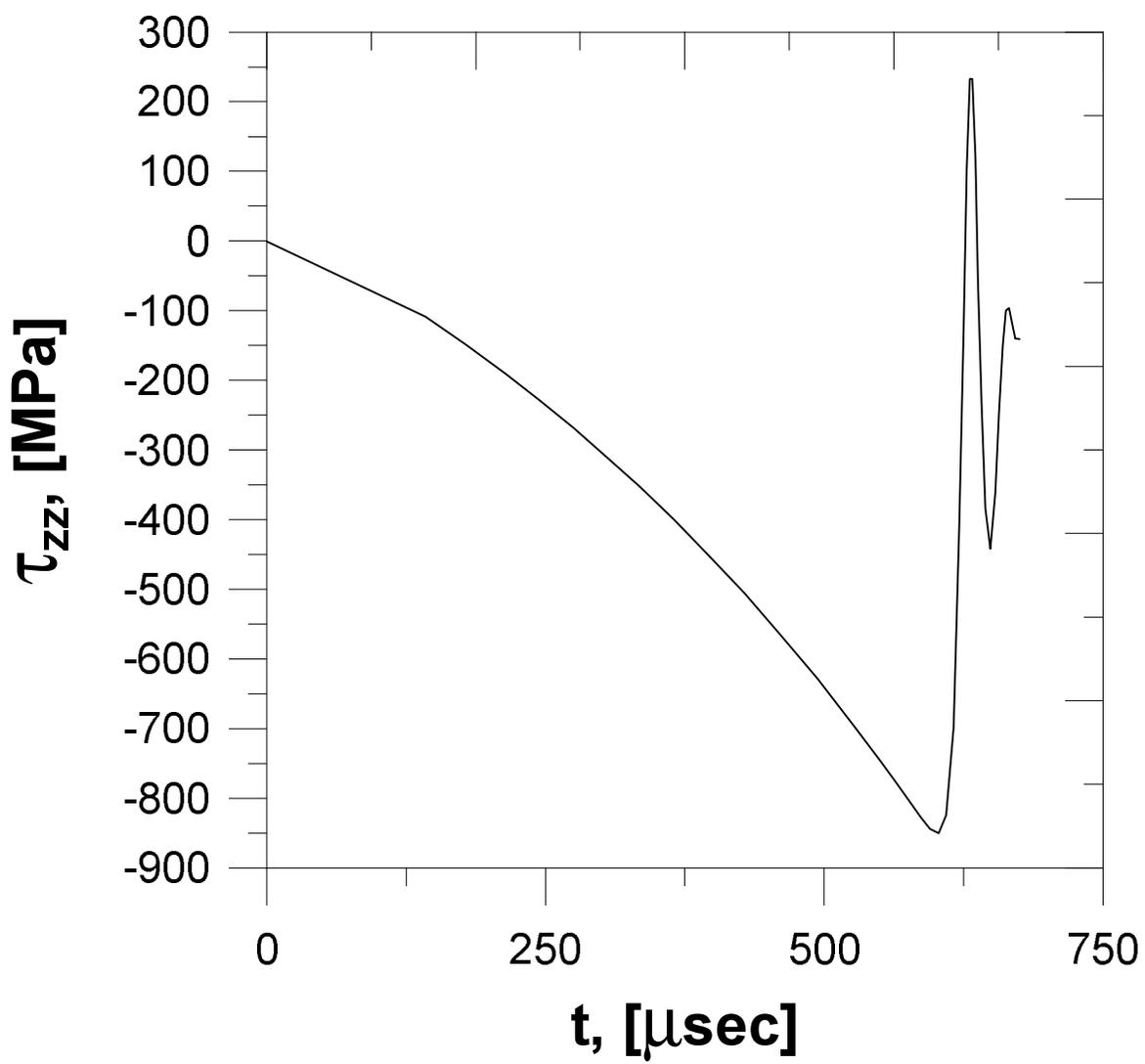
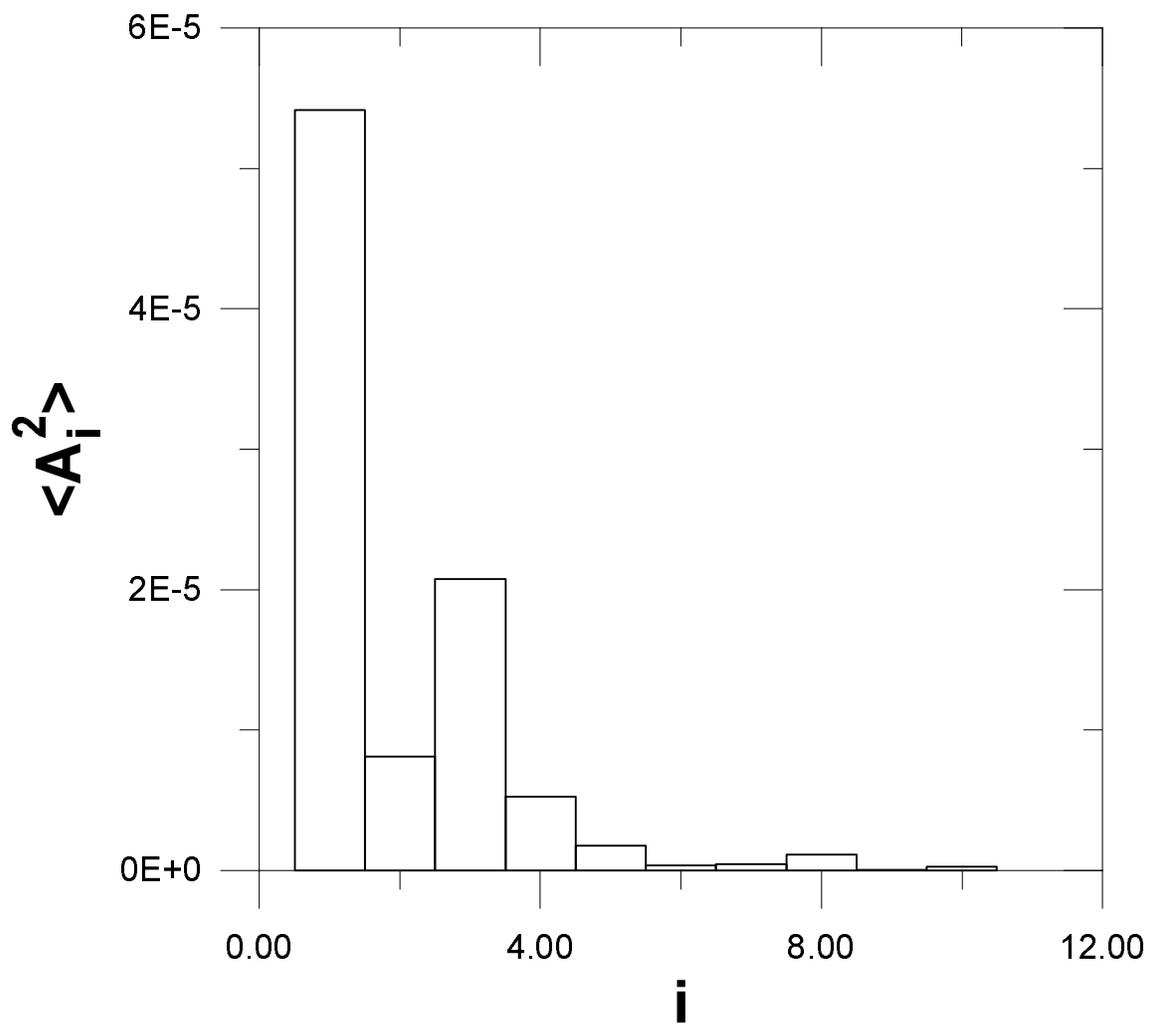
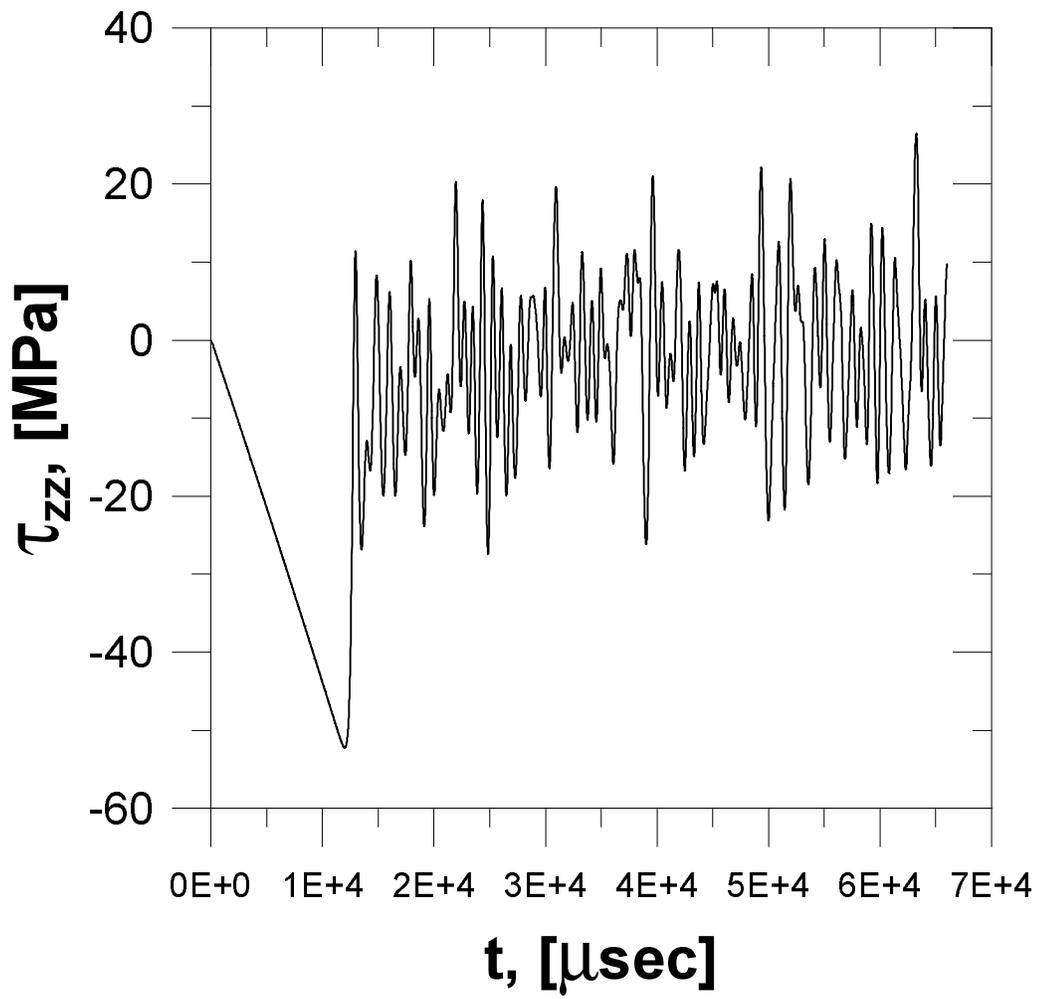
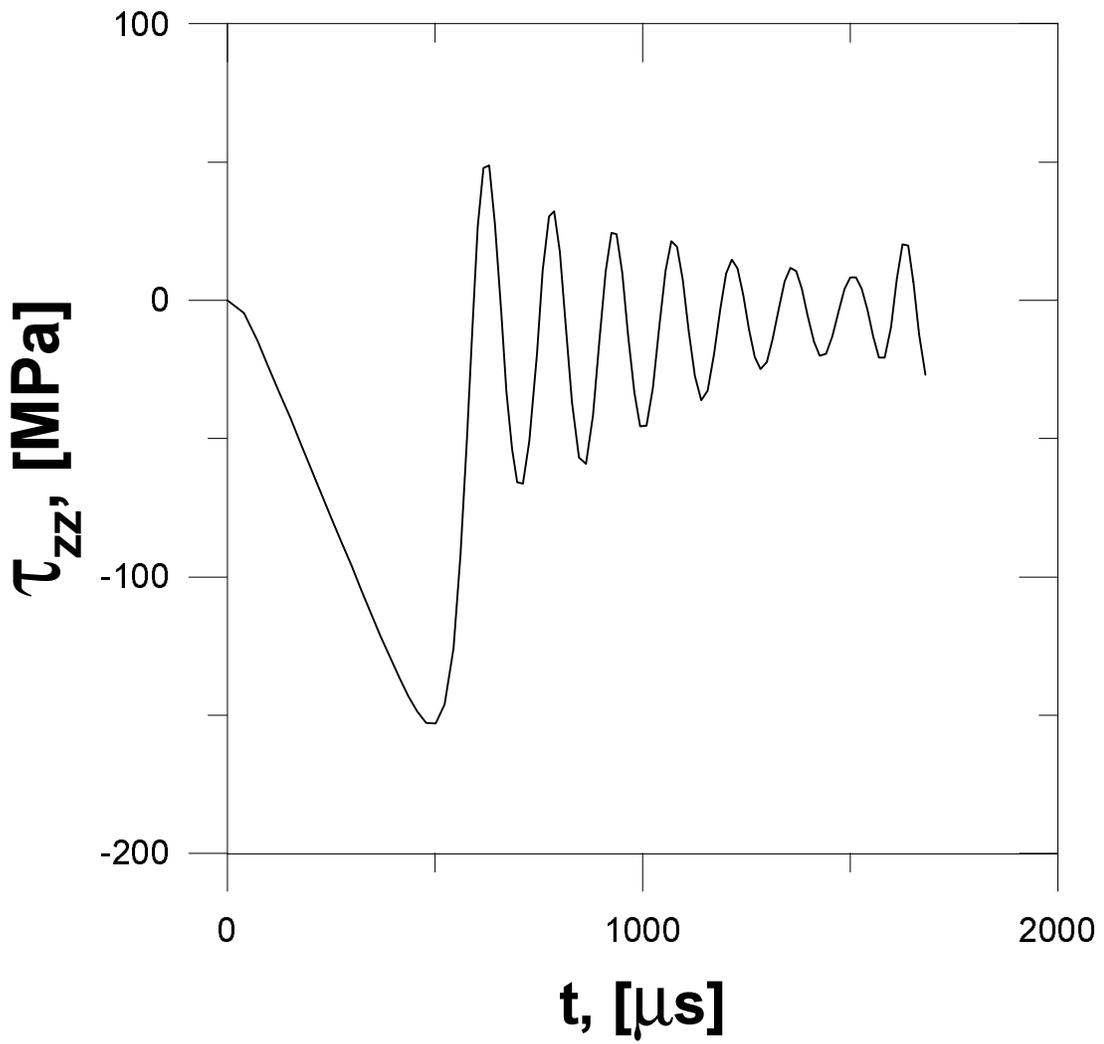
\end{document}